\documentclass[twocolumn,floats,floatfix,showpacs,amssymb,prd,superscriptaddress,nofootinbib]{revtex4-1}
\usepackage{graphicx, epsfig, amssymb} 
\usepackage{amsmath, amsfonts}
\usepackage{bm} 

\usepackage[linktocpage]{hyperref}
\usepackage[usenames]{color}

\def\be{\begin{equation}}
\def\ee{\end{equation}}
\def\beq{\begin{eqnarray}}
\def\eeq{\end{eqnarray}}

\begin{document}

\title{Superkicks in Ultrarelativistic Encounters of Spinning Black Holes}

\author{Ulrich Sperhake}
\affiliation{Institut de Ci\`encies de l'Espai (CSIC-IEEC), Facultat de Ci\`encies, Campus UAB, E-08193 Bellaterra, Spain.}
\affiliation{California Institute of Technology, Pasadena, CA 91109, USA.}
\affiliation{Department of Physics and Astronomy, The University of Mississippi, University, MS 38677, USA.}

\author{Emanuele Berti}
\affiliation{Department of Physics and Astronomy, The University of Mississippi, University, MS 38677, USA.}
\affiliation{California Institute of Technology, Pasadena, CA 91109, USA.}

\author{Vitor Cardoso}
\affiliation{CENTRA, Departamento de F\'{\i}sica, Instituto Superior T\'ecnico, Universidade T\'ecnica de Lisboa - UTL,
Av.~Rovisco Pais 1, 1049 Lisboa, Portugal.}
\affiliation{Department of Physics and Astronomy, The University of Mississippi, University, MS 38677, USA.}

\author{Frans Pretorius}
\affiliation{Department of Physics, Princeton University, Princeton, NJ 08544, USA.}

\author{Nicolas Yunes}
\affiliation{Department of Physics, Princeton University, Princeton, NJ 08544, USA.}
\affiliation{Department of Physics and MIT Kavli Institute, 77 Massachusetts Avenue, Cambridge, MA 02139, USA.}
\affiliation{Harvard-Smithsonian, Center for Astrophysics, 60 Garden St., Cambridge, MA 02138, USA.}

\date{\today}

\begin{abstract}
We study ultrarelativistic encounters of two spinning, equal-mass black holes
through simulations in full numerical relativity. Two initial data sequences
are studied in detail: one that leads to scattering and one that leads to a
grazing collision and merger. In all cases, the initial black hole spins lie
in the orbital plane, a configuration that leads to the so-called
{\emph{superkicks}}. In astrophysical, quasicircular inspirals, such kicks can
be as large as $\sim 3,000~{\rm km/s}$; here, we find configurations that
exceed $\sim 15,000~{\rm km/s}$. We find that the maximum recoil is to a good
approximation proportional to the total amount of energy radiated in
gravitational waves, but largely {\em independent} of whether a merger occurs
or not. This shows that the mechanism predominantly responsible for the
superkick is not related to merger dynamics. Rather, a consistent explanation
is that the ``bobbing'' motion of the orbit causes an asymmetric beaming of
the radiation produced by the in-plane orbital motion of the binary, and the
net asymmetry is balanced by a recoil. We use our results to formulate some
conjectures on the ultimate kick achievable in any black hole encounter.
\end{abstract}



\maketitle

\section{Introduction}
One of the more interesting consequences of binary coalescence is the
{\emph{recoil}} or {\emph{kick}} velocity that the center of mass can acquire
during the event. This possibility was first discussed by
Bekenstein~\cite{Bekenstein1973}.  Kicks are generated by an asymmetry in the
momentum carried away by gravitational waves (GWs): if more momentum is
carried away in any one direction, then the center of mass will ``react'' by
acquiring a velocity in the opposite direction to conserve momentum.

In one particularly interesting scenario where the black hole (BH) spins are
equal in magnitude, opposite in direction, yet within the orbital plane, the
recoil velocity can become quite large, a phenomenon that is sometimes called
a
{\emph{superkick}}~\cite{Campanelli2007,Bruegmann2007,Gonzalez2007a,Campanelli2007a}. At
first glance, it is somewhat surprising that this configuration can lead to
such a large recoil, as this is a highly symmetric orbit: the masses are
equal, the spins are anti-aligned, and the system's total angular momentum
equals the orbital angular momentum. Furthermore, the resultant kick velocity
depends sinusoidally on the initial phase of the binary, and linearly (at
leading order) on the magnitude of the individual BH spins.

A schematic explanation of the superkick was initially offered in
Ref.~\cite{Pretorius2007a}, as being due to the ``dragging of the inertial
frame'' of one BH relative to the other, and vice-versa.  This was expanded
upon in~\cite{2009PhRvD..80l4015K,Lovelace2009}, where it was pointed out
that, in addition to the frame-dragging effect, there is also a spin-curvature
coupling effect responsible for the super-kick at the same post-Newtonian
order.  From a distant observer's perspective, these effects cause the orbital
plane to ``bob'' up and down in a sinusoidal manner, while the binary
inspirals, with frequency equal to the orbital frequency. This bobbing motion
by itself does not directly produce the radiation that must be balanced by a
recoil. Rather, the bobbing causes the radiation produced by the binary's
in-plane orbital motion to be blue/red-shifted, in synchrony with the
bobbing. It is this asymmetry in the radiation pattern that ultimately results
in net linear momentum radiated in a direction orthogonal to the orbital
plane, balanced by the remnant BH moving in the opposite direction, after
coalescence.

More recently, however, Gralla et al. \cite{Gralla:2010xg} have argued, based
on an electromagnetic analogue model, that the bobbing motion is purely
``kinematical'' in nature, and not responsible for the recoil. Rather, they
speculate that the recoil has to arise in a process directly related to the
merger event which causes field momentum to be ``released'' and radiated to
infinity.  Here we present a first study of the ultrarelativistic scattering
of two BHs in the superkick configuration, in part to address the issue about
the origin of the superkick, and in part to continue our study of the
high-energy regime in BH collisions. Specifically, we study two families of
initial data, one leading to merger and one leading to scattering, although in
the latter the BHs interact strongly. In {\em both} cases we find essentially
{\em identical} recoil behavior of the center of mass following the
interaction: the recoil direction is orthogonal to the orbital plane, and the
magnitude varies sinusoidally with the initial phase, with a maximum
proportional to the {\em net} energy radiated in GWs. This is completely
consistent with the original scenario where bobbing-induced blue/red-shifting
of the radiated energy leads to the recoil. The only effect of the merger is
to slightly enhance the radiated energy, and hence the maximum recoil.

These conclusions do not necessarily imply that the electromagnetic analogue
in~\cite{Gralla:2010xg} is incorrect. However, since radiation-reaction
effects were not included in that study, it is conceivable that the same
phenomena would arise in the scattering of appropriately aligned magnetic
dipoles. In fact, then the two explanations described above might more be a
difference in semantics, i.~e.~an issue of whether one considers radiation to
be a ``release'' of ``field momentum'', which could happen regardless of
merger, as it does in the BH scattering case.

Several other interesting conclusions can be drawn from the results of this
study, other than implications for the nature of the mechanism of the
superkick. First, in the scattering cases, we also see situations where a
so-called {\em anti-kick} is present; i.e., where the maximum instantaneous
net linear momentum radiated is not equal to the final value. Thus, again,
explanations of this phenomenon relying on effects due on the presence of a
common horizon (as in Ref.~\cite{2010PhRvL.104v1101R}) cannot be the complete
picture. Second, to gauge the effect that spin in this configuration has on
the overall energy and {\em angular momentum} radiated in a merger, we compare
the results obtained for each sequence with those for an equivalent binary,
except the BHs initially have zero spin. We find that spin has very little
effect on the radiated energy and angular momentum. Third, we show that
subdominant effects in the spins scale as predicted by the ``spin expansion
formalism'' developed by Boyle, Kesden and Nissanke
\cite{Boyle2007a,Boyle2007b}. This is true for both merging and scattering
configurations; the latter result is nontrivial, since the spin expansion was
explicitly formulated for binaries that lead to mergers.  Fourth and last,
these simulations result in the largest superkicks seen to-date in merger
simulations, upwards of $\sim15,000{\rm km/s}$.  This is a factor of $5$
larger than the maximum yet seen in quasicircular BH coalescences, and $50 \%$
larger than those obtained by Healy et al. in hyperbolic
encounters~\cite{Healy2008}.  In Section \ref{speculations} we provide some
speculations on the maximum kick that could theoretically be achievable in any
ultrarelativistic encounter. We note that such enormous super-kicks are not
expected to occur in realistic astrophysical scenarios.

An outline of the rest of the paper is as follows.  Section~\ref{sec:
  numerics} discusses the numerical implementation of the problem and related
numerical uncertainties. In Section~\ref{sec: results} we present our results
on the radiated energy and linear momentum.  We conclude in Sec.~\ref{sec:
  conclusions} with a summary of our findings.  Throughout this work we use
geometrical units ($G = c = 1$), unless otherwise noted.

\section{Numerical simulations
\label{sec: numerics}}

We have performed numerical simulations with the {\sc Lean} code
\cite{Sperhake2006} which evolves the Einstein equations using the
Baumgarte-Shapiro-Shibata-Nakamura (BSSN) formulation
\cite{Shibata1995,Baumgarte1998} in combination with the moving puncture
method \cite{Baker2006,Campanelli2006}.  The exact form of our evolution
system is given by Eqs.~(11), (A1), (A4), (A6-A8) in Ref.~\cite{Sperhake2006}.
For evolving the shift $\beta^i$, we follow \cite{vanMeter2006} and employ a
first order in time version of the so-called ``Gamma driver'' [see their
  Eq.~(26)]. The free parameter $\eta$ is set to $\eta=0.7$ (in code units) in
all simulations. This corresponds to $M\eta=0.868$, where $M$ is the total
center-of-mass energy of the system.

The {\sc Lean} code is based on the {\sc Cactus} computational toolkit
\cite{Cactusweb} and uses mesh refinement provided by {\sc Carpet}
\cite{Carpetweb, Schnetter2004}. Initial data are calculated according to the
puncture method \cite{Brandt1997} with Bowen-York parameters \cite{Bowen1980}
using the {\sc Cactus} thorn {\sc TwoPunctures} based on Ansorg's spectral
solver \cite{Ansorg2004}. Apparent horizons are located and analyzed with
Thornburg's {\sc AHFinderDirect} \cite{Thornburg1996, Thornburg2004}.  GWs are
extracted using the Newman-Penrose scalar $\Psi_4$, as summarized in Appendix
C of Ref.~\cite{Sperhake2006}. The energy, linear and angular momentum carried
by GWs are obtained from $\Psi_4$ according to Eqs.~(2.8), (2.11) and (2.24)
of Ref.~\cite{Ruiz2007}.  For more details on the code we refer the reader to
Refs.~\cite{Sperhake2006, Sperhake2007}.

\subsection{Initial configurations}

All of our simulations are performed in the center-of-mass frame of the
binary, defined as the frame with zero ADM linear momentum~\cite{York1979}.
We determine the initial parameters of each BH under the assumption of
isolated horizons. This approximation is justified by the large initial
separations used for all simulations. We thus obtain the irreducible mass
$M_{{\rm irr},i}$ for BH $i=1$ and $2$ and calculate the BH rest mass from
Christodoulou's \cite{Christodoulou1970} relation
\begin{equation}
  M_i^2 = M_{{\rm irr},i}^2 + \frac{S_i^2}{4 M_{{\rm irr},i}^2},
      \label{eq: christodoulou}
\end{equation}
where $S_i$ is the spin magnitude of the $i$th BH.  The boost parameter is
defined by the ratio of dynamic to rest mass $\gamma = M_{\rm dyn} /
(M_1+M_2)$, where
\begin{equation}
  M_{\rm dyn}^2 = M_1^2 + P_1^2 + M_2^2 + P_2^2,
\end{equation}
and $P_i$ is the magnitude of either BH's initial linear momentum.  In this
work, we consider equal-mass binaries so that the boost of the individual BHs
equals $\gamma$.  In practice, both BHs start on the $x$-axis at location $\pm
x_0$ and their initial Bowen-York momenta are $\mathbf{P} = (\mp P_x,\,\pm
P_y,\,0)$, so that the initial orbital angular momentum is given by $L=d \;
P_y=2x_0 \, P_y$.

\begin{figure}[t]
  \centering
  \includegraphics[width=200pt,clip=true]{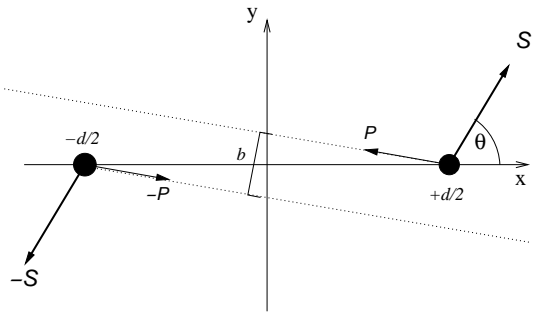}
  \caption{Illustration of the BH binary initial configuration.}
  \label{fig:config}
\end{figure}

With these definitions, we can characterize a binary initial configuration
using the following parameters: the boost parameter $\gamma$, the magnitude of
the dimensionless spin $\chi_i=S_i/M_i^2$ (where in all of our simulations
$\chi_1=\chi_2=\chi$), the initial separation $d$, the impact parameter
$b=L/P$ and the orientation of the spins measured by the angle $\theta$
relative to the coordinate axis connecting the initial BH positions (see
Fig.~\ref{fig:config}).
For both sequences, we fix the boost parameter $\gamma=1.52$, corresponding to
$P/M=0.374$, the dimensionless spin $\chi=0.621$ and the initial separation
$d=58.2~M$. The two sequences differ in the impact parameter; $b=3.34~M$ for the
s-sequence (scattering) and $b=3.25~M$ for the m-sequence (merging binaries). We
carried out a total of twenty simulations for selected values of the angle
$\theta$ in the range $[0^\circ,360^\circ]$.  For comparison, we also present
results from two nonspinning, equal-mass binaries with the same rest mass,
boost and impact parameters. Radiated energy and angular momenta, and (for the
merger cases) final horizon properties are summarized in Table \ref{tab:
  models} (some of these quantities have not yet been introduced, but they
will be defined later on in the paper).

\begin{table}
  \begin{ruledtabular}
  \begin{tabular}{l|cc|cc}
                    & {\bf Mergers} &                & {\bf Scatters} &\\
                    & Average       & Max. Dev.      & Average        & Max. Dev.      \\
    \hline
    $E_{\rm rad}/M$ & 0.295         & $2.3\%$        & 0.252          & $2.2\%$\\
    $E_{\rm phys}/M$& 0.265         & $2.6\%$        & 0.222          & $2.1\%$\\
    $J_{\rm rad}/J$ & 0.643         & $2.6\%$        & 0.580          & $1.2\%$\\
    $J_{\rm phys}$  & 0.605         & $5.2\%$        & 0.531          & $0.7\%$\\
    $M_{\rm irr}/M$ & 0.607         & $0.3\%$        & ---            & ---\\
    $j_{\rm fin}$   & 0.869         & $3.2\%$        & ---            & ---\\
    $j_{\rm QNM}$   & 0.890         & $4.4\%$        & ---            & ---\\
    $j_{\rm AH}$    & 0.889         & $2.2\%$        & ---            & ---\\
  \end{tabular}
  \end{ruledtabular}
  \begin{center}
  \caption{\label{tab: models}Initial and final parameters for the two
    sequences of binary models. Note that in all cases the estimated
    uncertainties in these quantities (not shown) from numerical truncation
    error or finite extraction radius is {\em larger} than the intrinsic
    variation within each sequence, {\em including} the two nonspinning
    comparison cases. Therefore, rather than list the values for all the
    separate cases, here we just list the average value, and the maximum
    deviation relative to the average. Note that for merger cases we only have
    apparent horizon information from roughly half the simulations, and so
    corresponding averages and deviations for the mass $M_{\rm irr}$ and spin
    $j_{\rm AH}$ only include those.}
    \end{center}
\end{table}

\subsection{Computational grid and uncertainties}

We have evolved all binary configurations on a numerical grid consisting of
ten nested refinement levels, three levels with one component centered on the
coordinate origin and seven levels with two components each, centered on
either BH. Using the notation of Sec.~II E of Ref.~\cite{Sperhake2006}, the
exact grid setup in units of $M$ (rounded to three significant digits) is
given by
\begin{eqnarray}
  && \left\{ (258,\,184,\,92) \nonumber \right. \\
  && \left. \quad \times (13.8,\,6.90,\,3.45,\,1.73,\,0.863,\,0.431,\,0.216),~h \right\}. \nonumber
\end{eqnarray}
Our standard resolution is $h=M/223$, but for convergence testing we have also
evolved one merger case using a coarser resolution $h_{\rm c}=M/195$ and
finer resolution $h_{\rm f}=M/250$. GWs have been extracted on a a set of six
concentric spheres of coordinate radii $R_{\rm ex}=57.5M$ to $86.3M$ in steps
of $5.76M$.

\begin{figure}[ht]
  \centering
  \includegraphics[width=240pt,clip=true]{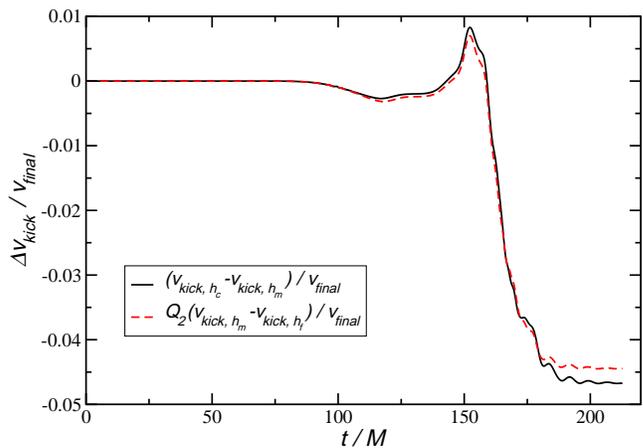}
  \caption{Convergence test for the recoil velocity $v_{\rm kick}(t) \equiv
    P_{\rm rad}(t) / M_{\rm fin}$.  The convergence factor $Q_2=1.459~$
    corresponds to second-order convergence.}
  \label{fig: conv}
\end{figure}

The convergence analysis for the recoil velocity is shown in Fig.~\ref{fig:
  conv}. Here we define a {\em time-dependent kick} as the quotient of the
radiated momentum and the final BH mass: $v_{\rm kick}=-P_{\rm rad}(t)/M_{\rm
  fin}$.  The figure demonstrates second-order convergence. Richardson
extrapolation reveals a relative uncertainty of the numerical kick velocity
obtained with medium resolution of about $9\%$.

\begin{figure}[t]
  \centering
  \includegraphics[width=240pt,clip=true]{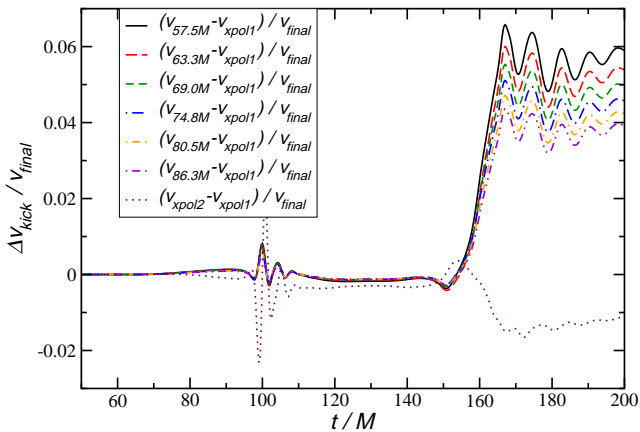}
  \caption{Difference of the recoil velocity $v_{\rm kick}(t) \equiv P(t)/M$
    at different extraction radii from the reference curve obtained by
    extrapolation according to Eq.~(\ref{eq: xpol1}), refered to as
    ``xpol1''. We also show the difference of a second fit assuming a
    quadratic term according to Eq.~(\ref{eq: xpol2}), referred to as
    ``xpol2''.}
  \label{fig: rex_vkick}
\end{figure}

A second main source of error is inherited from the use of finite extraction
radii. We study the resulting error by analyzing $v_{\rm kick}(t)$ extracted
for the high-resolution simulation of the test model at six different radii in
Fig.~\ref{fig: rex_vkick}. For this purpose we have first aligned the velocity
functions in time to compensate for differences in the propagation time, and
fitted the resulting curves with either of
\begin{eqnarray}
  v(t,r_{\rm ex}) &=& v_0(t) + \frac{v_1(t)}{r_{\rm ex}},
      \label{eq: xpol1} \\
  v(t,r_{\rm ex}) &=& v_0(t) + \frac{v_1(t)}{r_{\rm ex}}
                      + \frac{v_2(t)}{r_{\rm ex}^2}.
      \label{eq: xpol2}
\end{eqnarray}
The predicted recoil for infinite extraction radius is given by $v_0(t)$.
The fractional error in the velocity, as inferred from the difference between
the largest extraction radius used in practice ($r_{\rm ex}=86.3~M$) and the
extrapolated value, is roughly $4\%$.  We note that the main contributions to
the error in the velocity typically are opposite in sign: finite resolution
truncation error causes an underestimate of the recoil, while the use of
finite extraction radius results in an overestimation.

In the remainder of this paper we report radiated quantities
obtained at $r_{\rm ex}=86.3~M$ and at medium resolution and cite a
combined
error due to discretization and finite extraction radius of
$13\%$. For the reasons mentioned above, we consider this a rather
conservative estimate of the uncertainties.

In order to calculate the physical radiated momenta and energy (the quantities
with a subscript ``phys'' in Table~\ref{tab: models}), we exclude from the
extraction the early part of the gravitational waveforms up to $t-r_{\rm
  ex}=50~M$, which is dominated by spurious radiation due to the initial
data. For reference, the total radiated quantities that include the spurious
radiation are also shown in the table with a subscript ``rad''.

Before we discuss our results in more detail, we conclude this section with a
summary of further diagnostic quantities. The total center-of-mass energy of
the system is given by the ADM mass of the initial data as provided by the
spectral solver. The radiated momenta and energy enable us to calculate the
final BH mass
\begin{equation}
  M_{\rm fin} = M - E_{\rm rad}.
\end{equation}
In the case of scattering configurations this mass is to be interpreted as the
sum of the individual BH masses in the limit of large separation.  Balance
arguments further provide us with an estimate for the dimensionless spin of
the merged BH
\begin{equation}
  j_{\rm fin} = \frac{L-J_{\rm rad}}{M_{\rm fin}^2}.
\end{equation}
By virtue of the symmetry of the binaries studied in this work, the angular
momentum of the BH binary as well as that contained in the gravitational
radiation points in the $z$ direction, defined as the direction of the initial
orbital angular momentum.

We can also estimate the spin $j_{\rm QNM}$ of the final BH by fitting the
gravitational waveform at late-times with an exponentially damped
sinusoid. The (QNM) frequency and damping time of this signal can be inverted
to obtain $j_{\rm QNM}$ (see e.g. Refs.~\cite{Berti2006,Berti2009}).

An alternative measure for the final spin is given in terms of the irreducible
mass of the apparent horizon. For this purpose we rewrite Christodoulou's
relation (\ref{eq: christodoulou}) for the post-merger BH as
\begin{equation}
  j_{\rm AH} = 2\frac{M_{\rm irr}}{M_{\rm fin}}
       \sqrt{1- \frac{M_{\rm irr}^2}{M_{\rm fin}^2}}.
\end{equation}
%

\section{Results}
\label{sec: results}

In this section we describe the main results from our study. In
Sec.~\ref{sec_rad_e} we first discuss the total {\em energy} radiated, before
turning to the question of net {\em momentum} in Sec.~\ref{sec: recoil}. In
Sec.~\ref{sec_rad_p_e} we comment on the relationship between these
quantities, borrowing results from a wider set of published simulation
results, noting that in magnitude the ratio of these two quantities is nearly
constant.  Based on this observation, we describe several speculative
extrapolations to guess what the ``ultimate'' kick might be in
Sec.~\ref{sec_ult}. Finally, in Sec.~\ref{sec_anti} we comment on what our
results imply about the mechanism of the anti-kick.

\subsection{Radiated energy}\label{sec_rad_e}

Before we discuss in detail the radiation of linear momentum, we consider the
energy carried away in the form of gravitational radiation. Radiated energies
are listed in the rows labeled $E_{\rm rad}$ and $E_{\rm phys}$ in Table
\ref{tab: models}; again, the latter row excludes contributions due to spurious
radiation inherent in the initial data.

\begin{figure}[t]
  \centering
  \includegraphics[width=240pt,clip=true]{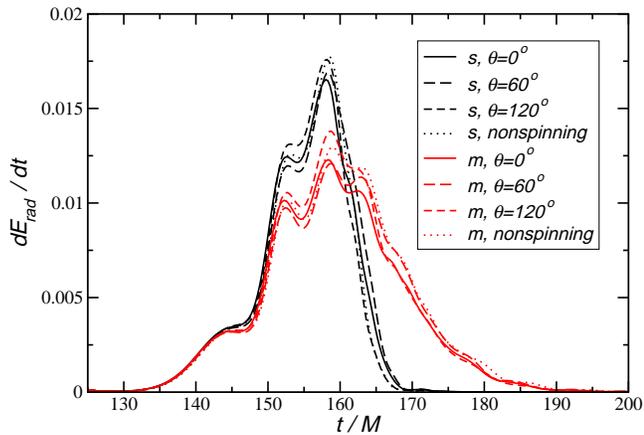}
  \caption{Radiated energy flux $dE_{\rm rad}/dt$ for a selected set of initial
  phase angles from both sequences ($s$ for scatter, $m$ for merger), and 
  including the nonspinning cases.}
  \label{fig: dEdt}
\end{figure}

Note that the radiated energy shows little variation (the ``maximum
deviation'' column) within either sequence. Also, the orientation of the spins
in the $xy$ plane has no impact on the outcome (merger or scattering) of the
binary interaction.  This confirms the observation made for the
astrophysically more relevant case of quasicircular superkick configurations,
as discussed for example in Sec.~III B of Ref.~\cite{Bruegmann2007}: the spin
orientation in the orbital plane does not significantly influence the dynamics
{\em within} the orbital plane.  Furthermore, the spin {\em magnitude} makes
little difference, as evidenced in that the maximum deviation listed includes
the nonspinning cases.  Fig.~\ref{fig: dEdt} shows the energy flux $dE_{\rm
  rad}/dt$ for a few cases from the two sequences.
The energy flux for all models within a sequence has similar levels of
agreement, so we restrict the number of curves in the figure for clarity.

In summary, the total radiated energy is essentially independent of the
orientation of the spins or, indeed, the presence of the spins in the first
place. 

\subsection{Gravitational recoil}
\label{sec: recoil}

The two sequences studied in this work either result in a merger or in a
scattering where no common apparent horizon forms and the BHs fly apart until
they can be regarded as isolated. For merging binaries the
total recoil is defined in the traditional manner: the linear momentum
radiated in the form of GWs has to be balanced by the recoil of the
post-merger BH. For scattering runs, we similarly define a {\em total kick} of
the binary system, but now the momentum due to the recoil is distributed over
two individual BHs instead of one. By virtue of the symmetry of
all configurations studied in this work, the two BHs 
acquire equal linear momentum after scattering, i.e. they move in the $z$ direction
with identical velocities.

\begin{figure}[t]
  \centering
  \includegraphics[width=240pt,clip=true]{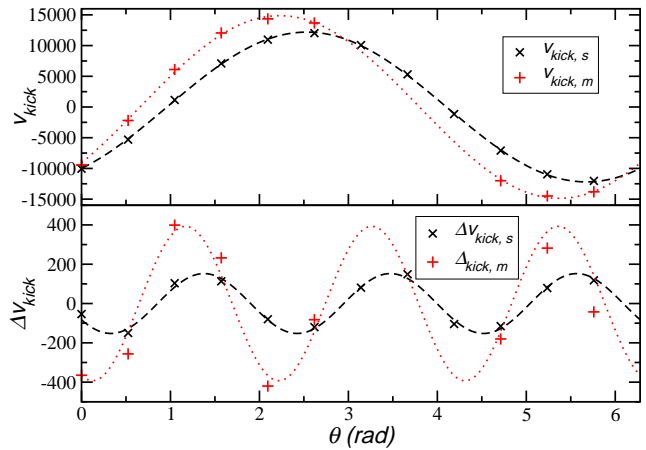}
  \caption{Top panel: the final recoil velocity as a function of the spin
    orientation $\theta$ for the m- and the s-sequence.  The curves are
    best-fit sinusoids to the data. Bottom panel: subdominant contribution to
    the final recoil velocity, obtained by subtracting from the total recoil
    the sinusoidal fit in Eq.~(\ref{dominant}).}
  \label{fig: vsm_cos}
\end{figure}

The final recoil velocity is shown in the top panel of Fig.~\ref{fig: vsm_cos}
for all parameters.  The data exhibit the same sinusoidal dependence of the
recoil on the orientation angle $\theta$ that was found for astrophysical
binaries. These best-fit sinusoids, also shown in the figure, are
\begin{eqnarray}
  \label{dominant}
  v_{\rm kick,\,s} &=& 12200~\cos(\theta - 2.53)~{\rm km/s}\,, \\
  v_{\rm kick,\,m} &=& 14900~\cos(\theta - 2.23)~{\rm km/s}\,.
\end{eqnarray}
Most significantly, the magnitude of the total radiated linear momentum is
quite similar between the two cases. It is also interesting to note, however,
that both the maximum recoil and the radiated energy are a bit larger for the
m-sequence than for the s-sequence. We will investigate this feature more
closely in the following section.

Boyle, Kesden and Nissanke \cite{Boyle2007a,Boyle2007b} systematically
expanded the mass, recoil velocity and spin of the remnant BH
resulting from a binary BH merger in terms of the binary's mass ratio
$q$ and of the individual spins $\chi_i$ of the two BHs. For
equal-mass ``superkick'' configurations where the two BHs have the
same Kerr spin parameter ($\chi_1=\chi_2=\chi$) their result for the final
kick can be expressed in the form
\begin{eqnarray}
\label{thirdord}
v_{\rm kick}&\simeq&
k_1\chi \cos(\theta-\theta_1) \nonumber \\
&+&k_3\chi^3 \cos(3\theta-\theta_3)+{\cal O}(\chi^5)\,,
\end{eqnarray}
The bottom panel of Fig.~\ref{fig: vsm_cos} shows that the subdominant
contribution to the kick is indeed well fitted by a function of this form for
{\it both} the merging and scattering sequences, whereas the spin expansion of
Refs.~\cite{Boyle2007a,Boyle2007b} considered mergers only.  A fit of the data
including third-order contributions in $\chi$ yields
\beq
&k^{(m)}_1=24119\,,\quad
k^{(m)}_3=-1722\,,\nonumber \\
&\theta^{(m)}_1=2.227\,,\quad
\theta^{(m)}_3=0.373
\eeq
for merging binaries, and
\beq
&k^{(s)}_1=19634\,,\quad
k^{(s)}_3=-636\,,\nonumber \\
&\theta^{(s)}_1=2.532\,,\quad
\theta^{(s)}_3=0.991
\eeq
in the scattering case. For comparison, the astrophysical mergers of
``superkick'' configurations lead to $k^{(J)}_1=3769$\,, $k^{(J)}_3=228$ for
the simulations by the Jena group \cite{Bruegmann2007} and to
$k^{(R)}_1=3622$\,, $k^{(R)}_3=216$ for the simulations by the Rochester group
\cite{Campanelli2007a} (cf. Table V of Ref.~\cite{Boyle2007b}). Considering
that these simulations were carried out by independent codes and considering
different values of the individual BH spins, these last sets of numbers are in
remarkable agreement and within the numerical errors.
%
\begin{table*}[ht]
  \begin{ruledtabular}
  \begin{tabular}{l|cccccc}
    Ref. &
    $\chi$ &
    $E_{\rm phys}/M$ &
    $v_{\rm max,1} ({\rm km/s})$ & 
    $v_{\rm max,3} ({\rm km/s})$ & 
    $M(v_{\rm max,1}/c) / E_{\rm phys}$ &
    $M(v_{\rm max,3}/c) / E_{\rm phys}$ \\
    \hline
    QC~\cite{Lousto2010a}  & $0.2\ldots 0.9$ & $0.0401$ & $3682$ & $3680$
                           & $0.306$ & $0.306$ \\
    H~\cite{Healy2008}     & $0.8$   & $0.128$  & $11988$ & $--$
                           & $0.312$ & $--$ \\
    s-seq.                 & $0.621$ & $0.220$& $19634$ & $19043$ 
                           & $0.298$ & $0.289$ \\
    m-seq.                 & $0.621$ & $0.265$& $24119$ & $22398$ 
                           & $0.303$ & $0.282$ \\
  \end{tabular}
  \end{ruledtabular}
  \caption{Summary of the initial dimensionless spin magnitude, radiated
    energy and maximum recoil velocity. For the latter quantity, we have two
    estimates, the first ($v_{\rm max,1}$) obtained using the leading order
    term in (\ref{thirdord}), the other ($v_{\rm max,3}$) also including the
    next-to-leading order term.  }
  \label{tab: skicks}
\end{table*}

Let us assume that the ``true'' values of the parameters for astrophysical
binaries are given by an average of the Rochester and Jena results:
$k_1=3695$\,, $k_3=222$. Then we find that, within the accuracy of the
numerics, the fitting coefficients of different orders in our relativistic
mergers and those in ``astrophysical'' mergers are roughly consistent with a
single proportionality relation: $|k^{(m)}_1/k_1|\simeq 6.5$ and
$|k^{(m)}_3/k_3|\simeq 7.8$.

As discussed below, these findings may have interesting implications to
estimate the maximum kick achievable in any binary BH encounter.

\subsection{\label{speculations}The relation between energy and recoil}\label{sec_rad_p_e}

Let us now investigate the relation between the radiated energy $E_{\rm phys}$
and the maximum\footnote{``Maximum'' in this context refers to variation over
  the orientation angle $\theta$ while keeping the center-of-mass energy
  constant. To avoid confusion, we will refer to the ``absolute maximum''
  achievable when we vary also the intrinsic parameters of the binary (center
  of mass energy, mass ratio and spin magnitudes) as the ``ultimate'' kick.}
recoil velocity $v_{\rm kick}$. We will use this relation in the next section
to extrapolate existing information on BH binaries and roughly estimate the
ultimate recoil achievable in any ultrarelativistic BH encounter.  

The search for a trend between radiated energy and recoil velocity requires
use of data from different configurations. We here consider data obtained by
Healy et al.  \cite{Healy2008} for hyperbolic encounters ($H$ in
Table~\ref{tab: skicks}) and the simulations of Lousto \& Zlochower
\cite{Lousto2010a} for approximately quasicircular binaries ($QC$ in
Table~\ref{tab: skicks}).  In order to assess the impact of subdominant
contributions to the expansion in Eq.~(\ref{thirdord}), we list in this table
two values for the maximum recoil velocity: (i) the leading order prediction
$v_{\rm max,1}$ using only the linear term and (ii) the value $v_{\rm max,3}$
obtained by also including the cubic term.  The table further shows the
radiated energy $E_{\rm phys}$ and the value (or range) of the spin magnitude
$\chi$ of the individual BHs considered. For either estimate of the maximum
kick velocity, we have calculated the ratio to the radiated energy.

Some comments on this table are in order.  First, in contrast with high-energy
collisions, BH binaries in quasicircular orbits radiate a significant fraction
of their energy during the early inspiral phase. For example, the total
radiated energy for a nonspinning equal-mass binary inspiral and merger is
approximately $0.05~M$~\cite{Scheel2008}, out of which about $70\%$
($0.035~M$) is radiated in the last two orbits; see
e.g.~\cite{Sperhake2006,Berti2007,Buonanno2007a,Baker2008a}. The radiated
energy for the quasicircular entry in Table \ref{tab: skicks} has been
obtained by averaging the values reported by Lousto \& Zlochower for their
$\chi=0.9$ sequence, but we cannot rule out that the relevant value may be
larger.  On the other hand, the early inspiral appears to have less impact on
the accumulated recoil, as intuitively expected: the GW flux increases more
slowly during early inspiral, so that the orbital average of the net momentum
flux is closer to zero (cf.~Sec.~III E of \cite{Bruegmann2007}). In the
remainder of this Section, we will employ the values listed in the table, but
one should keep in mind the above caveats and lack of high-precision data for
long quasicircular inspirals.

A second comment on Table~\ref{tab: skicks} concerns the study by Healy
et~al.~\cite{Healy2008}.  The sequence they simulated differs from the other
studies in that they vary the linear momentum of the BHs and, thus, the
kinetic energy of the binary.  Because of this, we only use one simulation
from their Figure~2, which provides a maximum kick. In units of the initial BH
rest mass $M_{\rm rest}$, they report an initial linear momentum of $P/M_{\rm
  rest}=0.308$, a radiated energy of $15\%$ and $v_{\rm kick} =9590~{\rm
  km/s}$ for this simulation. In order to use their data, we need to adjust
for the different normalization (with respect to the BH rest mass in
Ref.~\cite{Healy2008} and with respect to the total center-of-mass energy in
our comparison).  Including linear momentum, the dynamical BH mass is a factor
$1.174$ larger than the rest mass. Therefore we estimate the energy radiated
in their binary system as $12.8\%$ of the total energy of the system. Finally,
their data only allow us to estimate the maximum kick using the leading-order
extrapolation in the spin magnitude $\chi$.

In the final two columns of Table~\ref{tab: skicks} we show the ratio of the
maximum kick velocity, extrapolated to maximal spin $\chi=1$, to the radiated
energy per center-of-mass energy of the binary. For convenience, in this
column we measure the recoil velocity in units of the speed of light $c$. The
results suggest that, to leading order and for equal-mass binaries, the
maximum of the kick as we vary the orientation angle $\theta$ is proportional
to the radiated energy, which itself is to leading order independent of
$\theta$ and approximately equal to the energy radiated by the binary's
nonspinning counterpart.

\subsection{Conjectures on the ultimate kick}\label{sec_ult}

In this section, we speculate on several possible ways of extrapolating our
results to the ultimate recoil achievable in any binary black hole encounter
and on the related uncertainties.

Assuming that the scaling observed in the previous section remains valid for
arbitrary center-of-mass energies, we could derive a leading-order estimate of
the maximum kick possible in any BH binary encounter {\em if we knew the
  maximum radiated energy for equal-mass, nonspinning binaries}.  This maximum
energy has as yet not been determined, but ultrarelativistic grazing
collisions with boost factor $\gamma\approx 3$ have been found to radiate as
much as $35\pm5\%$ of the total energy of the system
\cite{Sperhake2009}. Extrapolation of the energy radiated by equal-mass
binaries has so far only been achieved for head-on collisions. In this case,
Ref.~\cite{Sperhake2008} predicts an increase in $E_{\rm phys}$ by a factor of
about $1.6$ as $\gamma$ increases from $3$ to infinity via
extrapolation. Based on this information, let us {\em assume} as a working
hypothesis, that grazing collisions of equal-mass, nonspinning BHs will result
in maximum energies up to about $50\%$ of the total mass. When combined with
the findings in Table \ref{tab: skicks}, this would result in an ``ultimate
kick'' of about $0.15$ times the speed of light, or $\sim 45,000~{\rm km/s}$.
In view of the assumptions made for this derivation, this prediction should
only be regarded as a rough estimate.

The most uncertain assumption in the previous analysis concerns the maximum
energy that can be radiated by equal-mass, nonspinning binaries. In fact, it
may well be possible that {\em all} the excess kinetic energy can be radiated
as we fine-tune the impact parameter around threshold.  Let us assume for the
sake of argument that this conjecture is true. Then, in the {\em
  large--$\gamma$ limit}, the maximum energy radiated would be $\sim 100\%$
for finely-tuned initial data. However, since spin does not change with boost,
most of this energy will be radiated when the spin is insignificant relative
to the total mass. The relevant question then becomes: {\em how much excess
  kinetic energy is left once the spin becomes significant?} Let us postulate
that, for a given $\gamma$, the kick is proportional to the relative excess
kinetic energy multiplied by an ``effective'' dimensionless spin $\chi_{\rm
  eff}$:
\begin{equation}
v \propto \frac{\gamma-1}{\gamma} \chi_{\rm eff}.
\end{equation}
If we were to let $\chi_{\rm eff}=\chi$, we would clearly have a problem, as
we would not be accounting for the fact that the spin becomes unimportant for
large $\gamma$, and we would arrive at a maximum kick speed at
$\gamma=\infty$. If instead we define $\chi_{\rm eff}$ as the spin angular
momentum normalized by the {\em total} mass, not the rest-mass: $\chi_{\rm
  eff}\equiv S/M^2 = \chi m^2/ (\gamma m)^2 = \chi/\gamma^2$, then
\begin{equation}
v = v_0 \frac{\gamma-1}{\gamma^3} \chi,
\end{equation}
where $v_0$ is the proportionality constant.  This scaling takes into account
the two limiting cases: for $\gamma=1$ no kinetic energy is radiated, so there
should be no kick; for $\gamma\to \infty$ the spin is irrelevant, so (again)
there should not be a kick. Quite interestingly, this ad-hoc ansatz yields a
maximum kick at $\gamma=3/2$, the case studied here, and hence we can use
these results to calculate $v_0$: this predicts a maximum kick of $\approx
24,000~{\rm km/s}$.
A higher $\gamma$ binary, sufficiently fine-tuned to the threshold of merger,
would presumably also yield a comparable maximum kick. For then it would loose
most of its excess kinetic energy to gravitational radiation, and hence the
last few orbits (that contribute most to the kick) prior to merger or scatter
will occur at a much lower $\gamma$.

A potentially more accurate way to estimate the maximum kick could employ the
Boyle-Kesden expansion, Eq.~(\ref{thirdord}). It is plausible to assume that
the coefficients $k_1$ and $k_3$ appearing in the fitting formula should only
be functions of $\gamma$.  The numerical estimate of these coefficients shows
that $|k^{(m)}_1/k_1|\simeq 6.5$ and $|k^{(m)}_3/k_3|\simeq 7.8$, so these
coefficients could have (to leading order) the {\it same} functional
dependence on $\gamma$: say, $k_1=\alpha k_3=f(\gamma)$, where $\alpha$ is a
constant. The main difficulty here consists of the fact that it is impossible
to determine this functional dependence having ``sampled'' the function at
only two values of $\gamma$ (i.e., $\gamma=1$ and $\gamma=1.52$). For example,
if we assume $f(\gamma)\sim \gamma^c$ we would get that $c \approx 4$ is a
good power-law scaling in the range $1\leq \gamma<2$. However it would be
perfectly legitimate to assume (say) that $f(\gamma)=a+b\gamma^c$, and then it
would be impossible to constrain the three parameters of the model with two
data points. As a matter of fact, we know that $f(\gamma)$ must asymptote to a
finite limit as $\gamma \rightarrow \infty$, because the kick velocity is
limited by the speed of light.

In summary, with presently available data, we cannot discriminate between the
scenarios discussed in this section: (i) a tentative guess of $45,000~{\rm
  km/s}$ if the proportionality between maximal kick and radiated energy holds
for arbitrary values of the boost parameter, (ii) a significantly lower
velocity of $\sim 24000~{\rm km/s}$ if the spin remains insignificant for the
binary dynamics until most of the kinetic energy has been radiated away; or
(iii) a more complex scenario where the ultimate kick can only be predicted by
a systematic expansion, as proposed by Boyle, Kesden and Nissanke.  Of course,
this uncertainty provides further motivation to simulate sequences of merging
binaries with larger $\gamma$ factors.

\subsection{Time dependence of $P_{\rm rad}$ and the ``anti-kick''}\label{sec_anti}

Up until now, we have exclusively analyzed the final value of the radiated
linear momentum $P_{rad}(t=\infty)$, but not its time-evolution.  Here, we
should distinguish between ``local'' BH dynamics and the presence of local
extrema in $P_{\rm rad}(t)$, as measured in the wave zone.  The former has
been studied, for example, in Refs.~\cite{2009PhRvD..80l4015K, Lovelace2009},
and reveals substantial reversals (of order $10^3~{\rm km/s}$) in local
estimates of the BH velocities. These large variations are not mirrored in the
radiated linear momentum as measured far away from the BHs\footnote{These
  large local velocities are consistent with the superkick explanation in
  Ref.~\cite{Pretorius:2007nq}, as such velocities are necessary to produce
  sufficient red/blue-shifts of the radiated momentum to account for the
  actual kick.}, and henceforth, with ``anti-kick'' we exclusively refer to
the deceleration visible in $P_{\rm rad}$ as measured far away from the BHs.

This anti-kick has been discussed in connection with ringdown radiation from
the remnant BH in Refs.~\cite{Schnittman2007a,LeTiec:2009yg,
  LeTiec:2009yf}. The addition of ringdown to post-Newtonian estimates of the
recoil generated by unequal-mass, nonspinning binaries has resulted in
excellent agreement with numerical calculations.
Ref.~\cite{2010PhRvL.104v1101R} attributed the anti-kick to deformations of
the common horizon after merger.

On the other hand, it is unclear if the anti-kick has any physical
significance at all. Consider the instantaneous momentum radiated in GWs in
any specific direction as given approximately by a sinusoidal function (from
orbital motion and then ringdown if a merger occurs), modulated by an envelope
proportional to the radiated energy, which is initially increasing (motion to
closest approach) and then decreasing (from ringdown in a merger scenario or
due to increasing separations in a scattering scenario). Then, there is no
{\em a priori} reason to expect that this function, integrated in time from
$t=-\infty$ to some $t=t_f$, should generically have its extremum at
$t_f=\infty$. As with many other properties related to kicks, the difference
between the maximum and the final value depends on the {\emph{phase}} of the
modulated sinusoid and it does not require any new physical mechanism to
explain it.

\begin{figure}[t]
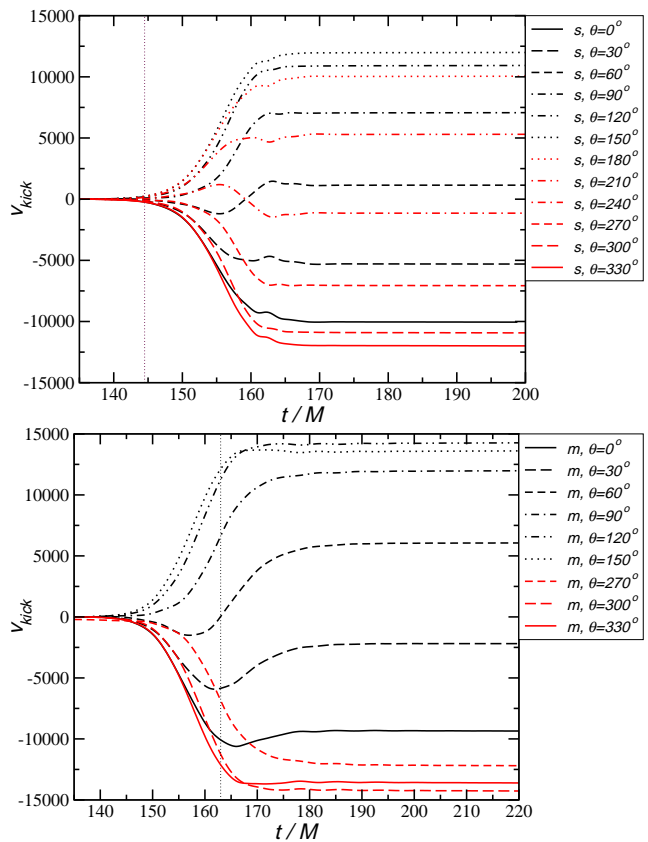

  \centering
  \includegraphics[width=240pt,clip=true]{vs.eps}
  \includegraphics[width=240pt,clip=true]{vm.eps}
  \caption{The recoil $v_{\rm kick}$ as a function of time extracted at
    $r_{\rm ex}=86.2~M$ for the s-sequence (upper panel) and the m-sequence
    (lower panel). For reference, the vertical lines mark the minimum
    coordinate separation of the punctures (for the scattering runs) and the
    common apparent horizon formation (for the mergers).  }
  \label{fig: vs}
\end{figure}

The one piece of evidence we can provide to the explanation of the anti-kick
can be seen in Fig.~\ref{fig: vs}, where we plot the radiated linear momentum,
converted into a recoil velocity by rescaling with the final mass of the
system, as a function of time. Anti-kicks (a decrease in the absolute
magnitude of the kick velocity as a function of time) are clearly present in
both merger and scattering cases. Hence, common horizon deformations cannot be
``the'' explanation, as a common horizon does not form in the scattering
cases.  Furthermore, since the magnitude of the anti-kick clearly depends on
the initial phase of the binary, the data is consistent with the
modulated-sinusoid description given in the previous paragraph.

%

\section{Conclusions}
\label{sec: conclusions}

In this work we have studied grazing collisions of equal-mass BH binaries with
antialigned spin angular momenta in the orbital plane -- the so-called
superkick configurations. We have studied two sequences with a moderate boost
of $\gamma\approx 1.5$: an m-sequence leading to formation of a common
apparent horizon; and an s-sequence where the two BHs eventually scatter off
to infinity. Within each sequence, we have varied the phase angle of the spin
angular momentum relative to the $x$-axis connecting the initial BH
positions. For comparison, we have also included two simulations of
equal-mass, nonspinning BHs with impact parameters corresponding to the two
sequences.

We have shown that the qualitative details of the recoil (sinusoidal
dependence on phase, and the maximum being proportional to the total energy
radiated) is independent of whether merger or scattering occurs. For the phase
angle resulting in the maximum recoil, we find $v_{\rm kick}=14,900~{\rm
  km/s}$ in the merger sequence, and $v_{\rm kick}=12,200~{\rm km/s}$ for the
scattering case.  Thus, the mechanism responsible for the superkick is, to
leading order, not related to merger dynamics. We have further found, as with
earlier observations for approximately quasicircular binary systems, that the
dynamics and net energy radiated in the orbital plane is essentially
unaffected by the presence or orientation of the spins.  We obtain radiated
energies of about $(22.2\pm2.2)\%$ and $(26.5\pm2.7)\%$ of the total
center-of-mass energy for the s- and m-sequence respectively, with no
dependence on the orientation of the spins to within the quoted uncertainties
of these numbers.  The corresponding nonspinning binaries also lead to similar
radiated energies to within the quoted uncertainties.

The ratio between the maximum recoil and the radiated energy for our two
sequences is very similar, and this has led us to compare our data to
available results from the literature on quasicircular binaries and hyperbolic
encounters of equal-mass, spinning binaries with opposite spins in the orbital
plane. Similar ratios between maximum radiated energy and recoil velocity are
found in all cases. Assuming that this scaling holds for arbitrary $\gamma$, a
{\emph{rough}} guesstimate for the ultimate kick would be around $45,000~{\rm
  km/s}$. On the other hand, it is equally possible that the spin of the
individual holes is insignificant until most of the kinetic energy has been
radiated away. We have estimated the resulting consequences by introducing an
{\em effective spin parameter}.  The ultimate kick would be smaller in this
case, about $24,000~{\rm km/s}$.  Even if neither of these simplifying
assumptions turn out to be valid, the determination of the ultimate kick can
be obtained from a systematic expansion of the kick dependence on the binary
parameters, as proposed by Boyle, Kesden and Nissanke \cite{Boyle2007a,
  Boyle2007b}.  In view of the substantial number of additional simulations
required for this purpose, we postpone such a study to future work.

Finally, we have analyzed the time evolution of the radiated linear momentum
with regard to the presence or absence of local extrema. The decrease in the
absolute magnitude of the recoil velocity after reaching a local extremum has
been dubbed ``anti-kick''. In both merging and scattering sequences, we
observe anti-kicks, though the particular value is dependent on the initial
phase.  Therefore, as with the superkick, anti-kicks are not a property
exclusively related to the formation and subsequent evolution of a common
horizon.

\begin{acknowledgments}
We thank Pablo Laguna for useful discussions. U.S. acknowledges support from
the Ram{\'o}n y Cajal Programme of the Ministry of Education and Science of
Spain, NSF grants PHY-0601459, PHY-0652995 and the Fairchild Foundation to
Caltech. E.B.'s research was supported by NSF grant PHY-0900735. V.C. is
supported by a ``Ci\^encia 2007'' research contract. FP and NY acknowledge
support from NSF grant PHY-0745779, and the Alfred P.Sloan foundation
(FP). This work was partially supported by the {\it DyBHo--256667} ERC
Starting Grant, by FCT -- Portugal through projects PTDC/FIS/098025/2008,
PTDC/FIS/098032/2008, PTDC/CTE-AST/098034/2008, CERN/FP/109306/2009,
CERN/FP/109290/2009, loni\_numrel05, an allocation through the TeraGrid
Advanced Support Program under grant PHY-090003 on NICS' kraken cluster and
the Centro de Supercomputacion de Galicia (CESGA, project number
ICTS-2009-40). Computations were partially performed on the Woodhen Cluster at
Princeton University. NY acknowledges support from the National Aeronautics
and Space Administration through Einstein Postdoctoral Fellowship Award Number
PF0-110080 issued by the Chandra X-ray Observatory Center, which is operated
by the Smithsonian Astrophysical Observatory for and on behalf of the National
Aeronautics Space Administration under contract NAS8-03060.  The authors
thankfully acknowledge the computer resources, technical expertise and
assistance provided by the Barcelona Supercomputing Centre---Centro Nacional
de Supercomputaci\'on.

\end{acknowledgments}

\end{document}